# Lattice specific heat of carbon nanotubes.

A. Sparavigna

*Dipartimento di Fisica, Politecnico di Torino*
*C.so Duca degli Abruzzi 24, 10129 Torino, Italy*

The lattice specific heat of carbon nanotubes is evaluated within the microscopic model proposed by Mahan and Jeon, published in the Physical Review B, in 2004. Phonons are considered for a single wall carbon nanotube in the armchair configuration. As expected, low temperature and high temperature regions show different behavior of specific heat. Carbon nanotubes are also displaying a very interesting lattice transport depending on the diameter, with a rather high thermal conductivity for small diameters.

## 1. INTRODUCTION

Carbon nanotubes are nanostructures with remarkable electronic and mechanical properties. As soon as they were discovered, their potential applications stimulated several theoretical and experimental studies. Electronic and lattice properties, such as specific heat and thermal transport coefficients, in carbon nanotubes [1-6] and nanowires [7] have been deeply investigated. Measurements with microfabricated devices show very interesting thermal transport properties in nanosystems [4]. Thermal conductivity increases as the tube's diameter decreases. At low temperatures, thermal conductance of carbon nanotube bundles follows the power law $T^\alpha$, where $T$ is the temperature, with an exponent $\alpha$ less than 2 (around 1.5). This fact suggested that the thermal transport in the bundles is like that of a quasi one-dimensional system. The same for specific heat: the measured specific heat differs from that of graphene and graphite, especially in the low temperature region, where quantization of the phonon band structure is observed.
The more evident property of thermal conductivity in carbon nanotubes is its increase as the carbon nanotube diameter descreases. A totally different behavior is observed in nanowires, where the thermal conductivity is reduced if the wire diameter is small. In nanowires, exponent $\alpha$ in the power law $T^\alpha$ ranges from 1 to 3, increasing with the wire section [7].

Here, in the framework of a recently proposed lattice model [8], we will discuss the thermal properties of the phonon system displayed by a single wall carbon nanotube in the armchair configuration. In Ref. [8], the authors deeply investigate the role of symmetry rules, pointing out that the violation of these rules destroys the flexural modes. They did not discuss the lattice thermal properties of the model. Here, the specific heat is estimated to see if the Mahan and Jeon model gives a behavior at low temperature in agreement with experimental data. A discussion on thermal conductivity follows.

## 2. PHONONS

A single wall carbon nanotube is a sheet of graphene rolled up along the line connecting two lattice points, into a seamless cylinder. According to the chosen line, a carbon tube can be achiral (zig-zag or armchair) or chiral. The seamless condition brings to continuity conditions on tube surface for the phonon vibrations and for the electron distribution. Several references discuss the behavior of electrons and electronic transport in nanotubes (for instance [9-11]) and phonon dispersions (see for instance [12,13]).

Here, to study the lattice thermal properties, we follow the lattice model proposed by G.D. Mahan and G.S. Jeon in Ref.[8]. These authors considered a "spring and mass" model with first and second neighbor bonds, and introduced a radial bond bending term in the potential, to include the curvature effect of the surface. The potential they propose satisfies the symmetry rules imposed by the cylindrical geometry. For the details of armchair nanotube lattice, see Figures 1 and 2 in Ref.[8].

An armchair $(n,n)$ tube with diameter R, has $n$ atoms A and $n$ atoms B along the circumference: each carbon atom A (or B) is connected with three nearest neighbor atoms B (A) at a distance $a$. Then, two integer numbers are required to describe the lattice sites $(l,m)$. $l$ is denoting the position of lattice sites in the direction of the nanotube axis and $m$, ranging from 0 to $n-1$, gives the positions of sites A and B on the circumference. Phonons in nanotubes have then two quantum numbers, one from the quantization of wave vectors in the direction of the tube axis, the other due to a quantization around the circumference.

Let us call $L$ the length of the tube, $N$ the total number of lattice points, $M$ the atomic mass and $\omega$ the angular frequency for a given wave vector $q_z$ along the axis direction, for given values of polarization $p$ and the quantum parameter $n$.

According to [8], phonon dispersions are evaluated. Oscillations of the lattice sites are traveling along the $z$-axis and, as found with the continuum models of hollow tubes, four acoustic modes are displayed. Two of them are the twist mode in tangential direction and the longitudinal stretching mode in the $z$-direction. The other two acoustic modes are the flexural oscillations. Flexural modes are also observed in wires [14]. The lower breathing mode in radial direction has a finite frequency for $q_z=0$, due to the lattice curvature, as observed in Ref. [13].

Three interactions, representing first neighbor, second neighbor and radial bond-bending forces, are used in this lattice model: their contributions are weighted with relative coefficients $r_j$ ($r_1 = 1.0$, $r_2 = 0.06$, $r_3 = 0.024$ respectively) [8]. Figure 1 shows the angular frequency for an armchair $(10,10)$ carbon nanotube, as a function of the reduced wavenumber $cq_z$, where parameter $c$ is equal to $\sqrt{3}a/2$. Here we use a scale factor $\Phi_o$ for the potential ($\Phi_o = 4 \times 10^{20}\ g\ cm^{-2} s^{-2}$), the value of which is chosen to give an acoustic longitudinal velocity of $16.7\ km/s$ and an acoustic twist velocity of $9.1\ km/s$ [8]. The lattice constant is $a = 1.42x10^{-8}\ cm$.

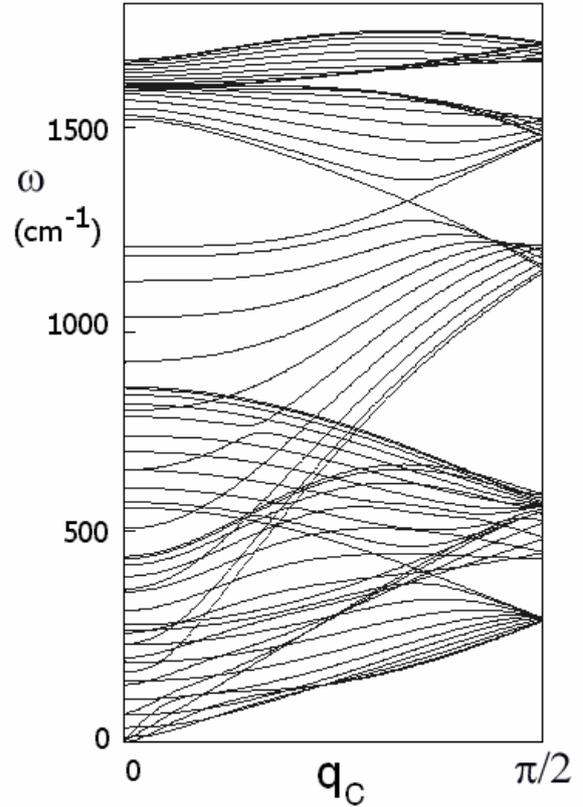

FIG.1: Phonon frequency $\omega$ as a function of wavenumber $q_z$ for an armchair tube (10,10). Dispersions are evaluated following Ref.[8].

## 2. LATTICE SPECIFIC HEAT

The lattice heat capacity is given by the following expression:

$$C = \frac{\hbar^2}{k_B T^2} \sum_{p,n} \frac{L}{2\pi} \int dq\, \omega^2 b_o (1+b_o) \quad (1)$$

and then the specific heat per atom is obtained, dividing by the number of the lattice sites. If we imagine a very long nanotube, the integral in Eq.(1) is a good substitute for the sum on all the values of the lattice wavenumber along the $z$-direction. Sum on polarization and eigenvalue number $n$ remains. $b_o$ is the unperturbed phonon distribution. Phonon frequency is according to Ref.[8].

Fig.2 shows the behavior of the specific heat as a function of the temperature for three armchair tubes, (5,5), (10,10) and (20,20), in comparison with the experimental data [1,6]. As observed by V.N. Popov [15], two distinct behaviors at low and high temperatures are clearly displayed by tube (5,5) and (10,10): they are due to phonon frequency quantization.

## 3. HEAT TRANSPORT

Phonon heat transport can be easily approached in the framework of the time relaxation approximation. The thermal conductivity due to phonons is linked to the perturbed phonon distribution $b$:

$$b - b_o = -\psi\, \frac{\partial b_o}{\partial(\hbar\omega)} \quad (2)$$

where $b_o$ is the equilibrium phonon distribution and $\psi$ a deviation function. The linearized Boltzmann equation for a solid subjected to a thermal gradient, written in the relaxation time approximation, turns out to be [16,17]:

$$k_B T \mathbf{v} \nabla T \frac{\partial b_o}{\partial T} = -\frac{1}{\tau}\psi\, b_o(1+b_o) \quad (3)$$

where $\mathbf{v}$ is the phonon group velocity and $\tau$ a phonon relaxation time. Eq.(3) is a very rough estimate of $\psi$.

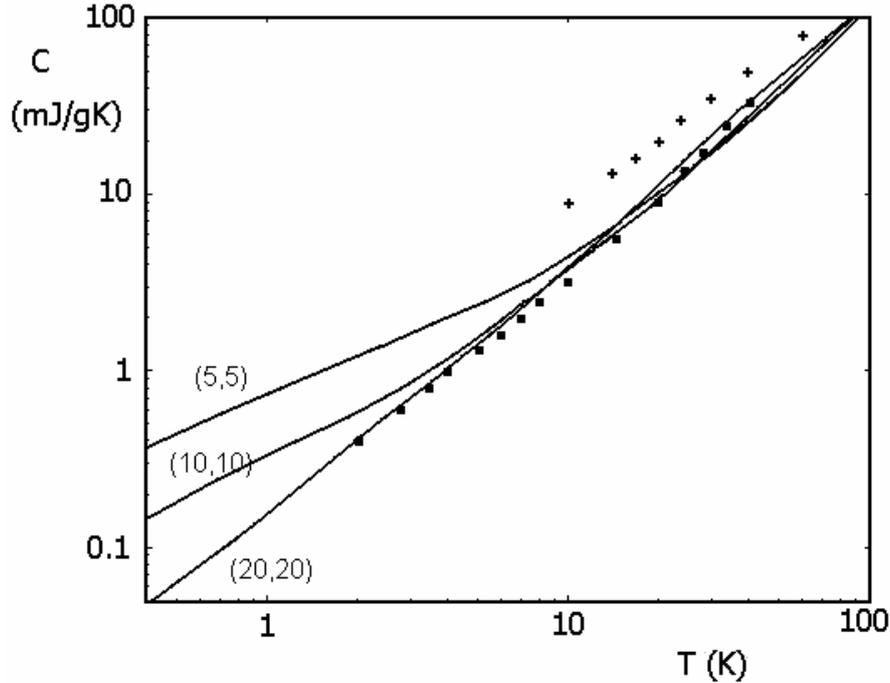

FIG.2: Specific heat as a function of temperature for single wall armchair nanotubes (5,5), (10,10), (20,20), compared with experimental data of Ref. [1] and [6].

The heat current density can be defined, for a nanotube with volume $\Omega = SL$ where $S$ is the section and $L$ the length, as:

$$U = -\frac{1}{2\pi S}\sum_{p,n}\int dq_z \, \hbar\omega \, v \, \psi \, \frac{\partial b_o}{\partial(\hbar\omega)} \quad (4)$$

where the sum on all the lattice wavenumbers along the $z$-direction is evaluated with an integral. A thermal current exists along the tube z-axis, if the nanotube is subjected to a gradient $\nabla T = \partial T/\partial z$. $v$ is the phonon velocity in the $z$-direction. The thermal conductivity is defined as that parameter $\kappa$ joining the heat current with the thermal gradient: $U = \kappa \, \partial T/\partial z$.

If the phonons are only subjected to boundary scattering, that is to the scattering occurring at the ends of the tube, the relaxation time is given by $\tau = L/v$ [17]. Thermal conductivity turns out to be:

$$\kappa = \frac{\hbar^2 a^3 L}{2\pi k_B T^2 S}\left(\frac{\Phi_o}{M}\right)^{3/2} \times \sum_{p,n}\int d\varsigma \, \overline{\omega}^{\,2} \, \overline{v} \, b_o(1+b_o) \quad (5)$$

where

$$\omega = a\sqrt{\frac{\Phi_o}{M}}\,\overline{\omega} \; ; \; \varsigma = cq_z; \; \overline{v} = |\partial\overline{\omega}/\partial\varsigma| \quad (6)$$

Eq.(5) is the kinetic model of thermal conductivity, if $\overline{v}$ is assumed equal to the mean value of the phonon velocity.

To compare with experimental data reported in Ref. 5 and obtained measuring the thermal conductivity of single carbon nanotubes, integer $n$ it put equal to 150: then the armchair tube diameter is 14 $nm$, as in experiments. To have an agreement with experimental data, the nanotube length $L$ must be equal to $6.2\,\mu m$. Thermal conductivity from Eq.(5) and experimental data are shown in Fig.3. The three curves in this figure are given for different values of tube diameter: thermal conductivity is strongly increased by reducing the diameter.

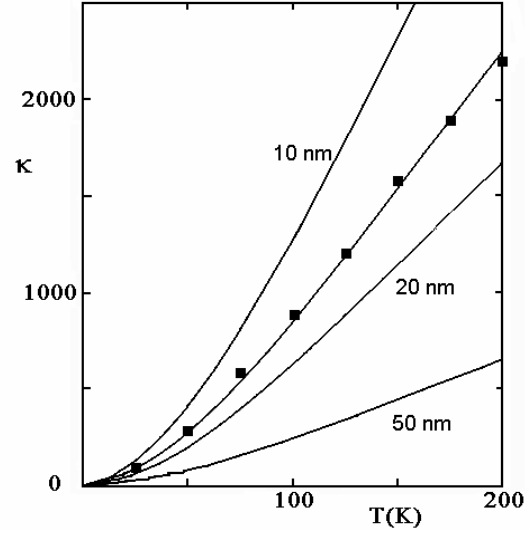

FIG.3: Thermal conductivity (in $W/m\cdot K$) as a function of temperature for a single wall nanotube with diameter 14 $nm$. Experimental data (squares) are from Ref. [5]. Curves refer to different values of the diameter.

Once more: this is only a very rough estimate of the thermal transport. A rigorous calculation of thermal conductivity must consider normal and umklapp three-phonon scatterings [18] and more details on phonon boundary scattering: both issues are very hard to discuss with a lattice model. Of course, MD simulations can give better performances in a more rigorous framework.

Curves in Fig.3 are following law $T^\alpha$ in the temperature range between 20 and 40 K, with $\alpha = 1.75$.

Other experimental data have been collected from bundles containing several nanotubes, all with parallel axes, in the hexagonal packing. Two bundles formed with tubes of different diameter (10 $nm$ and 148 $nm$) have the experimental outputs of Ref.[4]. In this reference, a behavior $T^\alpha$ with $\alpha = 1.5$, is deduced from the experimental points. Thermal conductivity is strongly reduced in the nanotube bundles, if compares with the values obtained in measurements on single nanotubes. With Eq.5, to obtain the experimental values of Ref.4, we must use a very short mean free path $L$ of about 25 $nm$.

Of course, if the nanutubes are packed in boundles, new phonon dispersions and phonon scattering mechanisms arise. Ref.6 shows how the hexagonal packing provides collective vibration modes. Moreover the nanotubes in the bundle can have different diameters, and this is another source of supplementary phonon scattering processes.

A last comment on the model of Mahan and Jeong concerns the four acoustic modes that it is displaying. Below 1 K, one-dimensional phonon transport is possible, in nanowires and nanotubes. At the low temperature regime, dominated by ballistic massless phonon modes, the phonon thermal conductance of a one-dimensional quantum wire is quantized, the quantum of thermal conductance being $\pi^2 k_B^2 T / (3h)$ where $h$ is Planck's constant [7,19].

Following the Landauer approach, it is possible to obtain the ballistic thermal conductance $G$, as in Ref.19 (the conductance $G$ obtained for a diffusive regime fails at very low temperature). The Landauer approach gives the thermal conductance as :

$$G = g \pi^2 k_B^2 T / (3h) \qquad (7)$$

where $g$ is the number of acoustic phonons. Acoustic modes are the channels through which the heat is transferred through a bridge (nanotube or nanowire) from a phonon reservoir to another. Any model of the brigde, nanotube or nanowire, must then involve no less than four acoustic modes.

## REFERENCES

1. W. Yi, L. Lu, A. Dian-lin, Z., Z.W. Pan, and S.S. Xie, Phys. Rev. B *59*, R9015, 1999
2. J. Hone, M. Whitney, C. Piskoti, and A. Zettl, Phys. Rev. B 59, R2514, 1999
3. J. Hone, I. Ellwood, M. Muno, A. Mizel, M.L. Cohen, A. Zettl, A.G. Rinzler, and R.E. Smalley, Phys. Rev. Lett. 80, 1042, 1998
4. L. Shi, D. Li, C. Yu, W. Yang, D. Kim, Z. Yao, P. Kim, and A. Majumdar, J. heat Transf. 125, 881, 2003
5. P. Kim, L. Shi, A. Majumdar, and P.L. McEuen, Phys. Rev. Lett. 87, 215502, 2001
6. A. Mizel, L.X. Benedict, M.L. Cohen, S.G. Louie, A. Zettl, N.K. Budraa, and W.P. Beyerman, Phys. Rev. B 60, 3264, 1999
7. D. Li, Y. Wu, P. Kim, L. Shi, P. Yang, and A. Majudar, Appl. Phys. Lett. 83, 2934, 2003
8. G.D. Mahan and Gun Sang Jeon, Phys. Rev. B, 70, 075405, 2004
9. M.S. Dresselhaus, G. Dresselhaus, and P.C. Eklund, Science of Fullerenes and Carbon Nanotubes, Academic Press, New York, 1996
10. S. Iijima, and M. Endo, Carbon, 33, 869, 1995
11. T. Ando, Semicond. Sci. Technol. 15, R13-27, 2000
12. V.N. Popov, V.E. Van Doren, and M. Balkanski, Phys. Rev. B 61, 3078, 2000
13. M.S. Dresselhaus, and P.C. Eklund, Adv. in Physics, 49, 705, 2000
14. N. Nishiguchi, Y. Ando, and M.N. Wybourne, J.Phys.: Condens. Matter, 9, 5751, 1997
15. V.N. Popov, Phys. Rev. B 66, 153408, 2002
16. J.M. Ziman, Electrons and Phonons: the Theory of Transport Phenomena in Solids, Clarendon, Oxford, 1960
17. G.P. Srivastava, The Physics of Phonons, Hilger, Bristol, 1990
18. A. Sparavigna, Phys. Rev. B 65, 064305, 2002; M. Omini, and A. Sparavigna, Nuovo Cim. D 19, 1537, 1997
19. L.G. C. Rego, and G. Kirczenow, Phys. Rev. Lett. 81, 235, 1998